# Widely tunable SNAP microresonators via translation of side-coupled optical fibers

**ISHA SHARMA[1],* AND MISHA SUMETSKY[1]**

[1]*Aston Institute of Photonic Technologies, Aston University, Birmingham B4 7ET, UK*
**i.sharma1@aston.ac.uk*

**We demonstrate free spectral range (FSR) tunable Surface Nanoscale Axial Photonics (SNAP) microresonators induced by side-coupled parallel optical fiber segments. By translating one segment relative to the other, we tune the coupling length from 900 μm to 100 μm and thereby tune the microresonator FSR from 5 pm to 50 pm, with an estimated precision of better than 0.003 pm. The microresonator Q-factor exceeds $10^5$ and can potentially be significantly increased in a clean lab environment. Possible applications of the demonstrated device include miniature and low-loss tunable delay lines and optical frequency comb generators, as well as ultraprecise tunable optical sensors.**

Optical microresonators are the fundamental building blocks of photonic devices due to their compact dimensions, high quality factors, and unique spectral characteristics [1-3]. The tunability of these optical microresonators is critically important in several applications, including cavity QED [4-6], optomechanics [7, 8], frequency microcomb generation [9, 10], and optical classical and quantum signal processing [11, 12]. Conventional tuning approaches, such as mechanical stretching, thermal tuning, and nonlinear optical effects, typically yield only minor changes in the free spectral range (FSR) of microresonators [13-19]. However, extensive tuning of the FSR of microresonators, which is of special importance for a range of the noted applications, remains challenging because it requires substantial changes to resonator dimensions or refractive index, which are difficult to achieve in compact monolithic designs.

Surface Nanoscale Axial Photonics (SNAP) platform offers a promising solution to this challenge. Commonly, SNAP microresonators are formed by a nanoscale effective radius variation (ERV) along optical fibers [20, 21]. The introduced ERV leads to localized changes in the fiber's cutoff wavelengths (CWLs), which govern the propagation of whispering-gallery modes (WGMs) near the fiber surface. This nanoscale ERV enables exceptional control over WGMs, which circulate around the fiber circumference and slowly propagate along the fiber axis. In particular, WGMs stop at turning points, where their wavelengths coincide with CWLs, and can be confined between two turning points along the fiber axis, forming a SNAP microresonator [20].

Several techniques for *tuning the axial FSR* of SNAP microresonators have been demonstrated. One of them is illustrated in Fig. 1(a). It employs *fiber stress*, introduced by fiber bending, which causes nonuniform variation in the fiber refractive index [22, 23]. The refractive index changes leading to the creation of a SNAP microresonator can also be introduced by *local heating* of an optical fiber [24] (Fig. 1(b)). In both methods, the introduced refractive-index variation has a relatively large millimeter-scale length along the fiber axis, since mechanical stress and heat distributions are difficult to tune at shorter lengths.

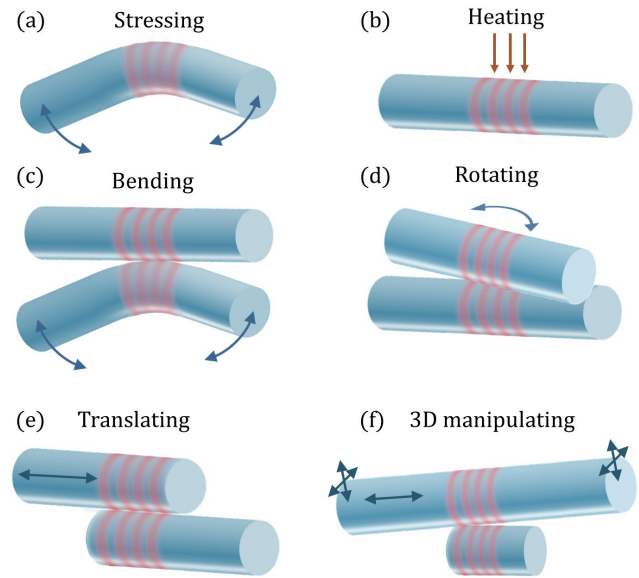

**Fig. 1.** Different methods for the introduction of reconfigurable SNAP microresonators by (a) stressing, (b) heating, (c) bending, (d) rotating, (e) translating, and (f) 3D manipulating of optical fibers.

Alternatively, Ref. [25] demonstrated that reconfigurable SNAP resonators can be created by *bending* an optical fiber side-coupled to another fiber (Fig. 1(c)). In this configuration, a SNAP resonator is induced by localized coupling between WGMs propagating in individual fibers rather than by the fiber stress. Advantageously, now the bent fiber curvature can be much smaller than that in a stress-induced resonator, while the resonator's axial length can be much smaller, as small as 100 μm [25]. However, miniaturizing this device while maintaining the same fiber curvature in the microresonator region (to preserve the FSR) requires a dramatic increase in the bending forces applied at the fiber ends, ultimately limiting the achievable degree of miniaturization and/or the FSR tuning range.

Recently, we demonstrated a different approach, illustrated in Fig. 1(d), in which a SNAP resonator is induced near the crossing point of two optical fibers and tuned by *mutual fiber rotation* [26]. In contrast to the approach of Ref. [25] (Fig. 1(c)), here the resonator tuning does not require the application of mechanical forces that increase with the tuning range and device miniaturization. It was shown in [26] that the millimeter-scale variation in the resonator's axial length corresponds to a dramatically small milliradian-order variation in the rotating angle. This property is advantageous for miniaturizing the proposed resonator system and integrating it with MEMS.

Building on the work described above, the present study demonstrates SNAP microresonators created by relative translation of side-coupled optical fibers, illustrated in Fig. 1(e), which enable controlled variation of the FSR. We will show below that, advancing beyond our previous rotating scheme [26], the developed approach achieves FSR tunability with a precision exceeding that of Ref. [26] by two orders of magnitude.

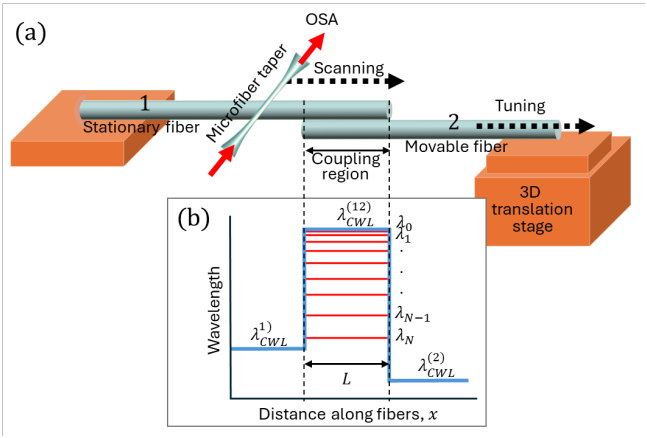

**Fig. 2.** (a) Schematics of the experimental setup. (b) Illustration of a rectangular microresonator $\lambda_{CWL}(x)$ formed at the side-coupled region of optical fibers.

Our experimental setup is illustrated in Fig. 2(a). To fabricate the proposed tunable microresonators, we used two silica fiber segments with radius $r_0 = 62.5$ μm, refractive index $n_0 = 1.44$, and carefully cleaved facets (Fig. 3). The fibers were cleaned in methanol, flame-treated, and then positioned parallel to each other and brought into side-coupling contact as illustrated in Fig. 2(a). Light was launched into fiber 1 (a stationary fiber) using a transverse input-output microfiber taper connected to Luna-5000 Optical Vector Analyzer (Luna OVA).

In the coupling region, the WGMs in fiber 1 and fiber 2 (the movable fiber) interact and couple to each other. The shifts of a CWL in the coupling region form a rectangular resonator illustrated in Fig. 2(b). The WGMs in this resonator propagate slowly along the coupling region and undergo reflection at the axial positions of the fiber ends. To experimentally analyze the effect of inter-fiber coupling on the behavior of CWLs and microresonators, the input-output microfiber was translated along fiber 1, periodically touching it at a spatial resolution of $2$ μm. The OVA measured the *spectrograms* of the transmission power $P(x, \lambda)$ as a function of cutoff wavelength $\lambda$ and microfiber position $x$ along the axis of the coupled fibers configuration.

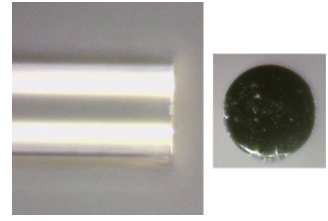

**Fig. 3**. The side and facet images of an optical fiber used in the experiment.

Figures 4(a)–(i) present the spectrograms of microresonators formed along the fiber coupling region (schematically depicted in the insets on the right side of each spectrogram), with the coupling length $L$ gradually reduced in $100$ μm increments from $900$ μm to $100$ μm. The overall shape of these resonators is defined by the CWL, $\lambda_{CWL}(x)$, where the left and right sides correspond to the CWL of the uncoupled sections of fiber 1 and fiber 2, respectively, and the central region reflects the CWL shift induced by fiber coupling. These CWLs correspond to WGMs with different azimuthal and radial quantum numbers. A positive CWL shift, $\Delta\lambda$ (approximately 0.2 nm in our experiments), within the coupling region gives rise to microresonators supporting localized WGM eigenstates. These microresonators can be tuned by varying the length of the side-coupled fiber segment. Sample transmission spectra, $P(x, \lambda)$, measured at specific positions $x$ (indicated by dashed vertical lines) are shown in Figs. 4(b), (d), (f), and (h). The largest Q-factor measured for these microresonators was $Q = 3 \cdot 10^5$. This moderate value is attributed to imperfections such as non-ideal fiber facet cleaving (which could be improved with tapered fiber ends), direct contact of the input-output microfiber waist with the fiber, and surface contamination in the coupling region in our uncontrolled lab environment. The FSR observed in the spectrograms varies from $\sim 5$ pm in Fig. 4(a) to $\sim 50$ pm in Fig. 4(h).

The slow axial propagation of the WGMs along the SNAP microresonator can be described by a one-dimensional wave equation having the form of the Schrödinger equation [20]. In this equation, the role of potential is played by CWL, $\lambda_{CWL}(x)$, illustrated in Fig. 2(b). Here, $\lambda_{CWL}(x) = \lambda_{CWL}^{(1)}$ and $\lambda_{CWL}^{(2)}$ in fibers 1 and 2, standing alone, $\lambda_{CWL}(x) = \lambda_{CWL}^{(12)}$ at their coupling region, and $\Delta\lambda = \lambda_{CWL}^{(12)} - \lambda_{CWL}^{(1)}$. In our experiment, we have $\Delta\lambda = 0.2$ nm (see spectrograms in Fig. 4). The rectangular quantum well model [27], applied within the framework of SNAP microresonator theory [20], shows excellent agreement with our experimental results. Specifically, the number of confined eigenstates in the quantum well is estimated for large $N$ from the WKB quantization rule as

$$N \cong \frac{L}{\Lambda}, \quad \Lambda = \frac{\left(\lambda_{CWL}^{(1)}\right)^{3/2}}{2^{3/2} n_0 \Delta\lambda^{1/2}}. \tag{1}$$

For our experimental parameters, $n_0 = 1.44$, $\lambda_{CWL}^{(1)} = 1.55$ μm, and $\Delta\lambda = 0.2$ nm, we find $\Lambda = 33.5$ μm. Thus, for example, we have $N = 18$ for $L = 600$ μm and $N = 6$ for $L = 200$ μm in exact agreement with spectrograms in Figs. 4(d) and (h), respectively.

The average axial FSR of our microresonators can be determined from Eq. (1) as $\Delta\lambda_{FSR} = \Delta\lambda/N = \Delta\lambda\Lambda/L$. For the same values of $\Delta\lambda$ and $\Lambda$ and $L = 500$ μm, we have $\Delta\lambda_{FSR} = $ 13 pm, in a reasonable agreement with our experimental data (Fig. 4(e)). From the latter expression for $\Delta\lambda_{FSR}$, the FSR tuning precision, $\delta\lambda$, can be expressed through the translation stage precision, $\delta L$, as $\delta\lambda = \Delta\lambda\Lambda L^{-2}\delta L$. For example, for the translation stage precision exceeding $\delta L = 0.1$ μm and a characteristic microresonator length of $L = 500$ μm, the achievable FSR tunability precision can exceed $\delta\lambda = 0.003$ pm. For comparison, at the same translation stage precision, this precision is two orders of magnitude better than that achievable in the rotating fiber setup we considered previously [26].

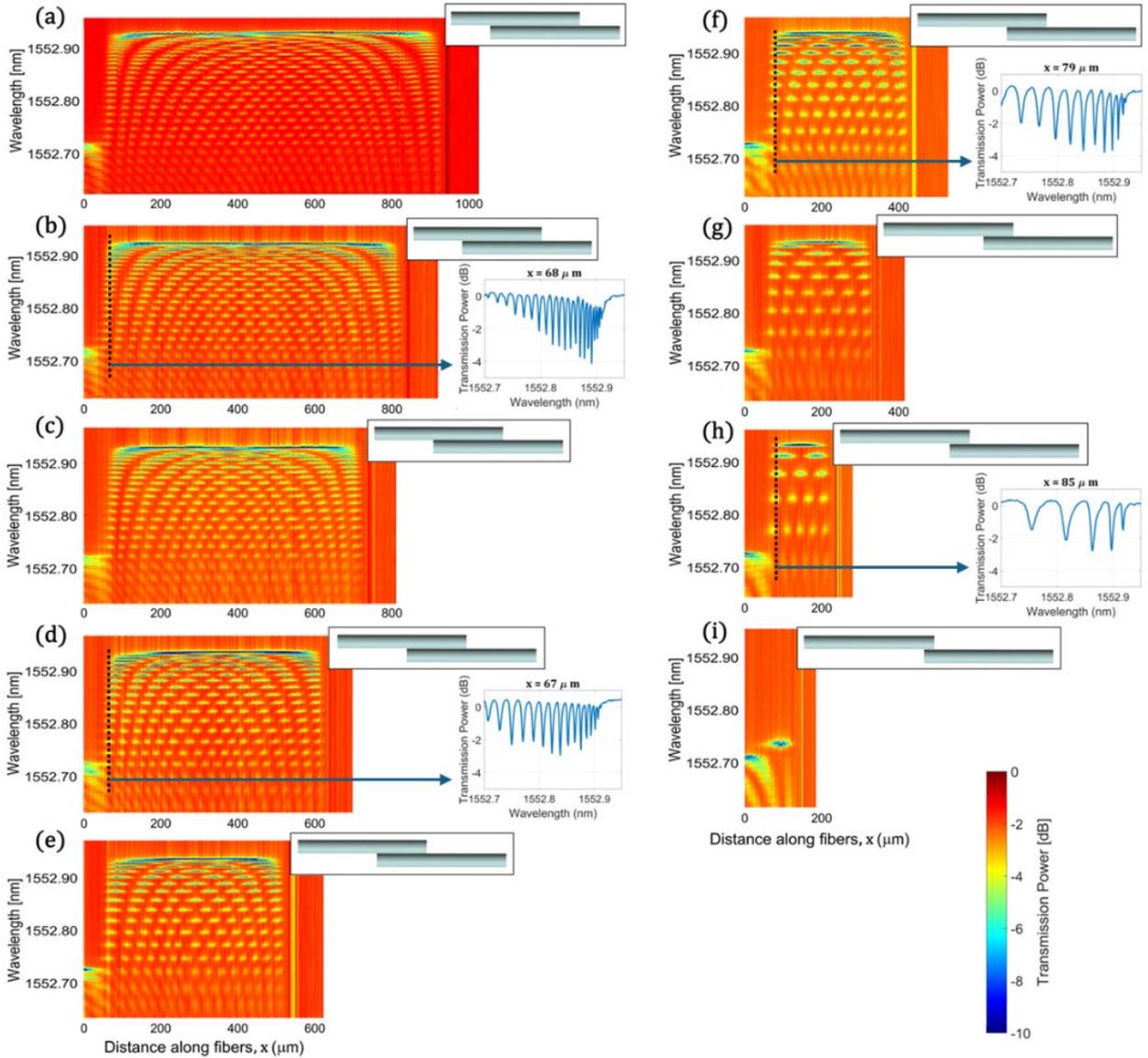

**Fig. 4.** (a)-(i) Spectrograms of microresonators $P(x, \lambda)$ induced along the fiber coupling region (illustrated in the insets at the left side of spectrograms) whose length was gradually reduced with 100 μm steps from 900 μm to 100 μm. In (b), (d), (f), and (h), we have added the spectra of these resonators at the positions $x$ indicated by dashed vertical lines.

In conclusion, we have experimentally demonstrated tunable SNAP microresonators formed at the side-coupled region of two parallel optical fiber segments, in which the axial FSR is controlled by relative fiber translation. By varying the coupling length from 900 μm to 100 μm, the FSR was continuously tuned from approximately 5 pm to 50 pm, in good agreement with the rectangular quantum-well model of SNAP resonators. The measured quality factor reached $Q \approx 3 \times 10^5$, limited primarily by non-ideal facet quality, direct microfiber contact, surface contamination, and the uncontrolled laboratory environment. With improved fiber facet preparation, controlled sub-micron gaps between the fibers and the input-output microfiber, and operation in a clean environment, the achievable Q-factor is expected to approach the $10^8$ level routinely demonstrated in other types of silica microresonators [2, 3, 15, 28]. In this regime, scattering losses would be dominated by residual nanoscale surface roughness rather than environmental contamination, enabling stable long-term operation and reproducible device performance.

One of the advantages of the translation-based approach compared to previously reported approaches [25, 26] is its high tuning precision, compatible with the order-of-magnitude FSR tunability. For a commonly achievable translation stage resolution of 0.1 μm and a resonator length of $L \sim 500$ μm, the estimated FSR tuning precision exceeds 0.003 pm, roughly two orders of magnitude superior to that of the rotating-fiber configuration [26].

The translating fiber configuration demonstrated here and the rotating configuration reported in Ref. [26] can be combined in an advanced fiber-manipulation platform enabling full three-dimensional control of optical fibers and, thereby, exploration of the advantages offered by all fiber degrees of freedom, as illustrated in Fig. 1(f). We suggest that such a device could be interfaced with integrated photonic circuits [29] to realize record compact and low-loss fiber-based tunable delay lines [30], optical frequency comb generators, signal-processing elements, and optical sensors generalizing their previously demonstrated stationary versions [31-33].

**Funding.** Engineering and Physical Sciences Research Council (EPSRC) grants EP/W002868/1 and EP/X03772X/1, Leverhulme Trust grant RPG-2022-014.

**Disclosures.** The authors declare no conflicts of interest.

**Data availability.** Data underlying the results presented in this paper may be obtained from the corresponding author upon reasonable request.

**References**
1. R. K. Chang, A.J. Campillo (Eds.), *Optical Processes in Microcavities* (World Scientific, Singapore, 1996).
2. K. J. Vahala, "Optical microcavities," Nature **424**, 839 (2003).
3. A.B. Matsko (Ed), *Practical Applications of Microresonators in Optics and Photonics* (CRC Press, 2009).
4. W. Von Klitzing, R. Long, V. S. Ilchenko, J. Hare, and V. Lefèvre-Seguin, "Frequency tuning of the whispering-gallery modes of silica microspheres for cavity quantum electrodynamics and spectroscopy," Opt. Lett. **26**, 166 (2001).
5. Y. Louyer, D. Meschede, and A. Rauschenbeutel, "Tunable whispering-gallery-mode resonators for cavity quantum electrodynamics." Phys. Rev. A **72**, 031801 (2005).
6. H. Pfeifer, L. Ratschbacher, J. Gallego, C. Saavedra, A. Faßbender, A. von Haaren, W. Alt, S. Hofferberth, M. Köhl, S. Linden, and D. Meschede, "Achievements and perspectives of optical fiber Fabry–Perot cavities," Appl. Phys. B **128**, 29 (2022).
7. T. J. Kippenberg and K. J. Vahala, "Cavity opto-mechanics," Opt. Express **15**, 17172 (2007).
8. I. Favero and K. Karrai, "Optomechanics of deformable optical cavities," Nat. Photon. **3**, 201 (2009).
9. L. Chang, S. T. Liu, and J. E. Bowers, "Integrated optical frequency comb technologies," Nat. Photon. **16**, 95 (2022).
10. Y. Sun, J. Wu, M. Tan, X. Xu, Y. Li, R. Morandotti, A. Mitchell, and D. J. Moss, "Applications of optical microcombs," Adv. Opt. Photon. **15**, 86 (2023).
11. C. Lian, C. Vagionas, T. Alexoudi, N. Pleros, N. Youngblood, and C. Ríos, "Photonic (computational) memories: tunable nanophotonics for data storage and computing," Nanophotonics **11**, 3823 (2022).
12. J. H. Ko, Y. J. Yoo, Y. Lee, H. H. Jeong, and Y. M. Song, "A review of tunable photonics: Optically active materials and applications from visible to terahertz," iScience **25**, 104727 (2022).
13. A. A. Savchenkov, V. S. Ilchenko, A. B. Matsko, and L. Maleki, "Tunable filter based on whispering gallery modes," Electron. Lett. **39**, 389 (2003).
14. D. Armani, B. Min, A. Martin, and K. J. Vahala, "Electrical thermo-optic tuning of ultrahigh-Q microtoroid resonators," Appl. Phys. Lett. **85**, 5439 (2004).
15. M. Pöllinger, D. O'Shea, F. Warken, and A. Rauschenbeutel, "Ultrahigh-Q tunable whispering-gallery-mode microresonator," Phys. Rev. Lett. **103**, 053901 (2009).
16. M. Sumetsky, Y. Dulashko, and R. S. Windeler, "Super free spectral range tunable optical microbubble resonator," Opt. Lett. **35**, 1866 (2010).
17. W. Jin, R. G. Polcawich, P. A. Morton, J. E. Bowers, "Piezoelectrically tuned silicon nitride ring resonator," Opt. Express **26**, 3174 (2018).
18. A. Kovach, J. He, P. J. G. Saris, D Chen, and A. M. Armani, "Optically tunable microresonator using an azobenzene monolayer," AIP Adv. **10**, 045117 (2020).
19. X. Jiang and L. Yang, "Optothermal dynamics in whispering-gallery microresonators," Light: Sci. & Appl., **9**, 24 (2020).
20. M. Sumetsky, "Theory of SNAP devices: basic equations and comparison with the experiment," Opt. Express, **20**, 22537, (2012).
21. M. Sumetsky, "Nanophotonics of optical fibers," Nanophotonics, **2**, 393 (2013).
22. D. Bochek, N. Toropov, I. Vatnik, D. Churkin, and M. Sumetsky, "SNAP microresonators introduced by strong bending of optical fibers," Opt. Lett. **44**, 3218 (2019).
23. M. Crespo-Ballesteros and M. Sumetsky, "Ultra-precise, sub-picometer tunable free spectral range in a parabolic microresonator induced by optical fiber bending," Opt. Lett. **49**, 4354 (2024).
24. A. Dmitriev, N. Toropov, and M. Sumetsky, "Transient reconfigurable subangstrom-precise photonic circuits at the optical fiber surface," In *2015 IEEE Photonics Conference*, Postdeadline paper, pp. 1-2. (2015).
25. V. Vassiliev and M. Sumetsky, "High Q-factor reconfigurable microresonators induced in side-coupled optical fibres," Light: Sci. & Appl. **12**, 197 (2023).

26. I. Sharma, and M. Sumetsky, "Widely FSR tunable high Q-factor microresonators formed at the intersection of straight optical fibers," Optica **12**, 890 (2025).
27. S. Flügge, *Practical quantum mechanics* (Springer Science & Business Media, 2012).
28. Z.-H. Zhou, C.-L. Zou, Y. Chen, Z. Shen, G.-C. Guo, and C.-H. Dong, "Broadband tuning of the optical and mechanical modes in hollow bottle-like microresonators," Opt. Express **25**, 4046 (2017).
29. D. Farnesi, S. Pelli, S. Soria, G. Nunzi Conti, X. Le Roux, M. Montesinos Ballester, L. Vivien, P. Cheben, and C. Alonso-Ramos, "Metamaterial engineered silicon photonic coupler for whispering gallery mode microsphere and disk resonators," Optica **8**, 1511 (2021).
30. M. Crespo-Ballesteros, and M. Sumetsky, "Miniature Tunable Slow Light Delay Line at the Bent Silica Fiber Surface," In *CLEO: Science and Innovations,* SS176_3 (2025).
31. M. Sumetsky, "Delay of light in an optical bottle resonator with nanoscale radius variation: dispersionless, broadband, and low-loss," Phys. Rev. Lett. **111**, 163901 (2013).
32. N. Toropov, S. Zaki, T. Vartanyan, and M. Sumetsky, "Microresonator devices lithographically introduced at the optical fiber surface," Opt. Lett. **46**, 1784 (2021).
33. X. Jin, X. Xu, H. Gao, K. Wang, H. Xia, and L. Yu, "Controllable two-dimensional Kerr and Raman–Kerr frequency combs in microbottle resonators with selectable dispersion," Photon. Res. **9**, 171 (2021).